\documentclass{pasj00}

\SetRunningHead{H. Negoro and S. Mineshige}
{Log-Normal Distributions in Cygnus X-1}

\Received{2002 April 30}
\Accepted{2002 July 19}
\Published{}

\begin{document}
\setcounter{page}{1}

\title{Log-Normal Distributions in Cygnus X-1:
Possible Physical Link with Gamma-Ray Bursts and Blazars}

\author{Hitoshi \textsc{Negoro}}
\affil{Cosmic Radiation Laboratory, RIKEN,
	2-1 Hirosawa, Wako, Saitama 351-0198
	}
\email{negoro@crab.riken.go.jp}
\and
\author{Shin \textsc{Mineshige}}
\affil{Yukawa Institute for Theoretical Physics, Kyoto
University, Sakyo-ku, Kyoto 606-8502}
\email{minesige@yukawa.kyoto-u.ac.jp}

\abst{
Prompted by recent discoveries of log-normal distributions in
gamma-ray/X-ray temporal variabilities of gamma-ray bursts (GRBs) 
and blazars,
we re-examine the X-ray variability of Cygnus X-1 in the hard/low state
using Ginga data in 1990.
It was previously reported that the distributions of the time intervals
between X-ray shots (flares) deviated from the Poisson distributions at short
time intervals: the occurrence of shots tended to be suppressed for
several seconds before and/or after a shot event.
Detailed analyses show that 
this deviation is larger for shots with larger peaks, and that 
the time-interval distributions for large shots approach
log-normal distributions with a peak at the interval of 7--8 s.
Furthermore, we also show that the peak-intensity distribution 
for the shots is consistent with the log-normal distribution, 
though no typical peak intensity can be seen.
This might indicate the presence of a physical link connecting the 
physics of black hole accretion flow and that of jet/GRB formation.
}
\kword{accretion, accretion disks --- methods: statistical ---
galaxies: jets --- gamma rays: bursts  --- X-rays: individual (Cyg X-1)}

\maketitle

\section{Introduction}
Gamma-ray bursts (GRBs) may be the most enigmatic objects
in present-day astronomy.
Among their peculiar properties, highly aperiodic time variability is 
undoubtfully one important aspect.
Among various timing analyses, we pay special attention to one by
Li and Fenimore (1996), who found log-normal distributions for
the peak fluence and peak time intervals of GRBs (for careful and detailed 
analyses, see Nakar, Piran 2002; Quilligan et al. 2002).
Although its physical meaning is not obvious, it is expected that
this feature may contain an important clue to understanding 
the central engine of GRBs.

Apparently similar, highly aperiodic intensity fluctuations are known in
X-rays from Galactic black hole candidates (GBHCs)
in the hard/low state,
though, in the other states, similar, highly aperiodic variations have not
been observed, except for rather periodic flares known as 
quasi-periodic oscillations (e.g., van der Klis 1995).
One may thus ask if the variability properties of GBHCs in the hard state
share common characters with GRBs.  An affirmative answer
is actually anticipated in a sense,  since the central engines 
of GRBs are often discussed in terms of accretion models 
(e.g., Narayan et al. 1992, 2001).
The basic idea underlying these models is that, 
whatever the origin might be (black hole--neutron star mergers, 
black hole--stellar core mergers, and so on),
the final configuration may likely be a stellar-mass black hole 
surrounded by a massive accretion disk or torus 
with a mass ranging between 0.01--1.0 $M_\odot$.
If so, it is natural to expect some similarities to exist
in gamma-ray and X-ray variabilities of GRBs and GBHCs.
This expectation becomes even more strengthened by the recent discovery
of log-normal distributions in the blazar variability 
by D. Yonetoku and T. Murakami (in preparation).

However, negative results have already been reported 
based on Ginga observations of the prime GBHC, Cygnus X-1
(Negoro et al. 1995, hereafter N95).
The time intervals between adjacent flare-like events (called X-ray shots),
$\Delta t$,
basically follow exponential distributions
with no peak, as expected from random (Poisson) 
distributions.
They also found that the peak intensities were exponentially distributed.

Yet, the entire story is not over.
At the same time, they discovered an interesting 
suppression of the shot occurrence at short time intervals ($\Delta t <$ 5--8 s).
The duration of the suppression, ``waiting time'', is longer for shots 
with a larger peak intensity.
Using RXTE data, Focke (1998) also found that a peak interval 
distribution could not be represented by a simple exponential function,
but by a broken exponential function.
In this Letter,
we put forward this discovery by performing new analyses 
of shots with large peaks using the same Ginga data of Cyg X-1 with 
N95, and discuss its physical meaning.

\section{Data and Peak Detection Algorithm}

We reanalyzed the Ginga LAC data of Cyg X-1 in the hard state on 1990 
May 9--11.
The minimum time resolution and the energy band 
of the data used were 7.8 (1/128) ms and 1.2--58.4 keV binned into
12 energy channels.
The net exposure time was 16.22 ks (N95).

The properties of time variations of Cyg X-1 
are basically different 
from those of GRBs, though both exhibit flare-like events,
so-called shots and bursts, respectively.
In addition to a big difference in the duration between 
extreme transient outbursts (lasting for less than $\sim100$ s in GRBs)
and the hard state (lasting for more than 1 month in BHCs),
the short-term variations on the timescales of less than 10 s
are also different. 
It is now clear that the variability of Cyg X-1 consists of
multi-components: shots with the typical peak duration of 0.1--0.2 s,
and spikes with much shorter duration and smaller peaks
(Negoro et al. 2001).
Furthermore, the shot peak intensities 
are rather smoothly distributed down to smaller ones, 
although the minimum ends are not clear due to the statistical 
count fluctuations (N95; Focke 1998).
These and the statistical fluctuations 
(worse S/N ratios) prevent us from defining a `reliable' peak 
for small shots and an `apparent' bottom between adjacent shots. 

Thus, we do not adopt the Li and Fenimore's (1996) peak detection algorithm,
but the same detection algorithm as N95.
This leads to interesting 
results, as shown later.
We changed, however, the algorithm to calculate the local
mean number of counts, $\langle{p}\rangle$, in order to select the shot peaks as free as 
possible from detection criteria:
we calculate $\langle{p}\rangle$, not from the numbers of counts at intervals 
of 32 s (N95), but from those in a bin to be checked and neighboring bins 
in $T_{\rm m}$ (including the checked bin) on its either side (Focke 1998). 
Namely, $\langle{p}\rangle$ in an $i$-th bin is estimated as 
$ \langle{p}\rangle (i)  = \sum_{j = i - T_{\rm m}/t_{\rm b} +1 }
	^{i + T_{\rm m}/t_{\rm b} -1} x_{j}$,
where $t_{\rm b}$ is the bin width, and $x_{j}$ is the number of 
X-ray counts in a $j$-th bin. 
This estimation gives a more definite local mean for each bin than 
the previous one,
and does not yield discontinuity at every 32 s intervals, as before.
We have confirmed, however, that this change
essentially has little effect on this and previous results.
We calculate $\langle{p}\rangle$ for each bin, and select peaks using
following criteria: the peak number of counts $p$ of the shot should be 
larger than 1.5--3 times $\langle{p}\rangle$, and should have the maximum number 
of counts within a certain duration $T_{\rm p}$ on either side.

We note that the parameter $T_{\rm m}$ has an influence 
on the peak interval distributions. 
However, this dependence is not recognized in distributions 
for shots with $p \gtrsim 2.2\langle{p}\rangle$, to which we pay attention here.
$T_{\rm m}$ is set to be 64 s, much longer than the duration which
we are interested in and shorter than the timescale of the long-term 
variations.
$T_{\rm p}$, on the other hand, affects the results of the peak intensity 
distribution, and is set to 0.25 s from the typical shot 
duration obtained from the mean profile of the shots (N95).
All of the data are binned on ($t_{\rm b} =$) 31.25 ms intervals 
in order to avoid large statistical count fluctuations. 
This also does not affect the following results qualitatively.
Thus, $\langle{p}\rangle$ is estimated from 
127.96875 s (=$2T_{\rm m} - t_{\rm b}$) data with a checked bin 
being at the center.

\begin{figure}[t]
 \begin{center}
    \FigureFile(76mm,150mm){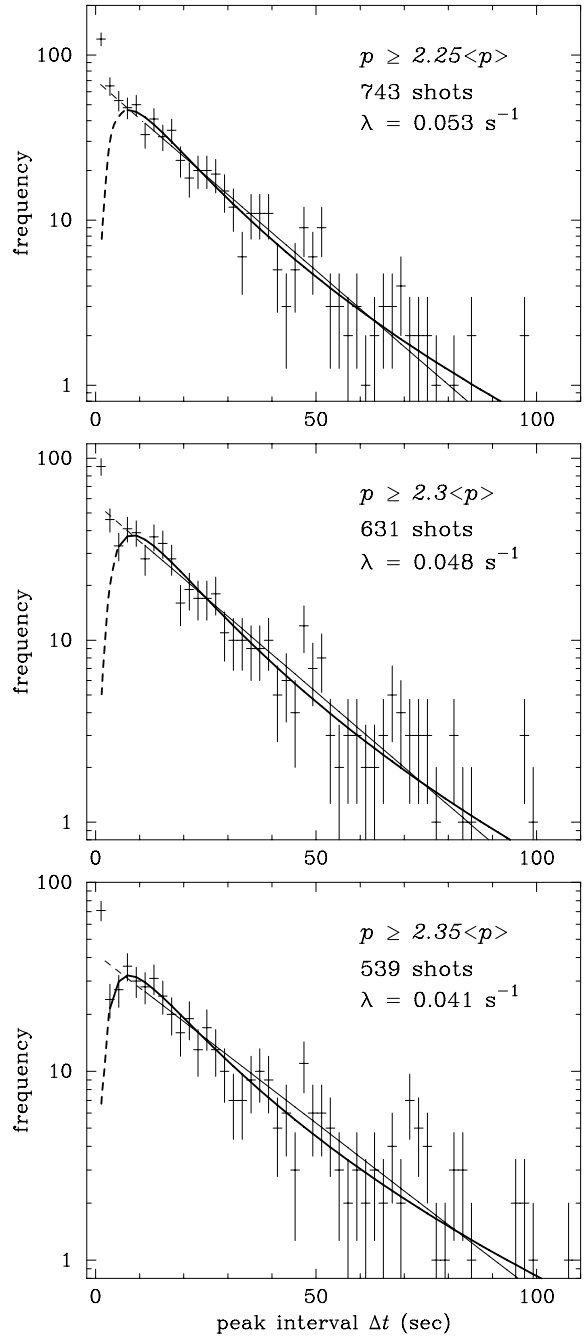}
 \end{center}
\caption{
Peak interval distributions around a critical peak intensity 
criterion (see text).
The thin and thick lines are the best-fit Poisson and log-normal 
distributions, respectively. 
The dashed lines are those extrapolations to short intervals (unfitted regions).
A peak intensity criterion, the number of accumulated shots,
and an occurrence rate obtained by the exponential fit at 
$\Delta t > 10.25$ s are shown in each panel.
}
\end{figure}

\section{Distributions of Peak Intervals and Intensities}

A log-normal function in this paper is expressed as
\begin{equation}
   \begin{array}{cc}
	f(x) = \frac{\displaystyle 1}{\displaystyle \sqrt{2\pi}\sigma} 
		\exp [ - \frac{\displaystyle (\log x - \log \mu)^2}
			{\displaystyle 2\sigma^2} ]
	& \ \ \ (x > 0),
   \end{array}
\end{equation}
and the best-fit parameters and errors are obtained by
maximum likelihood fits using 
{\tt QDP} in {\tt FTOOLS} version 5.0.4.

\subsection{Peak Interval Distribution}

\begin{figure}[t]
   \FigureFile(80mm,60mm){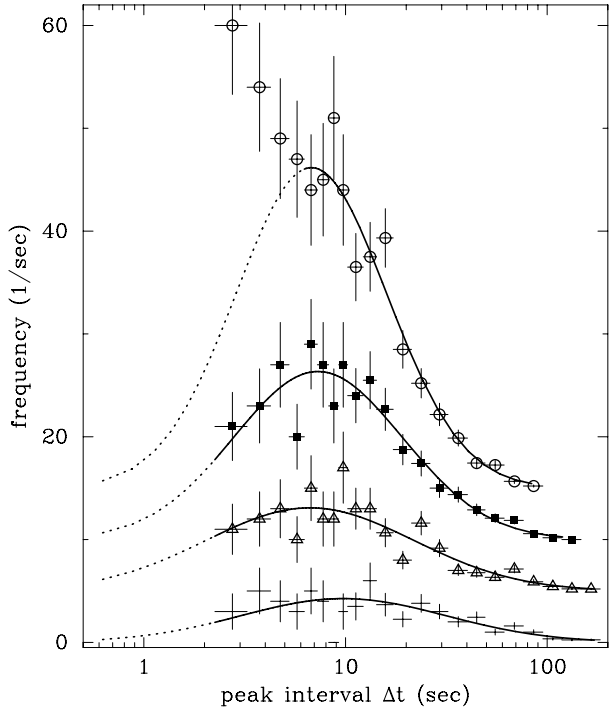}
\caption{
Peak interval distributions obtained by various peak-intensity criteria,
the best-fit log-normal distributions (solid lines), and 
those extrapolations (dotted lines).
The upper 3 distributions are shown shifted up 5, 10, 15 frequencies
for visual clarity.
Peak criteria $p/\langle{p}\rangle$ from the top to the bottom are 2.2, 2.35, 2.5 
and 2.6, and the numbers of accumulated shots 
are 871, 539, 353 and 266, respectively.
Note that no significant dependence of the log-normal peak positions
$\mu$  on the magnitude of the shots is recognized.
}
\end{figure}

 Peak (time-)interval distributions for shots can generally be described by
an exponential function with deviations at short intervals: enhancement at 
$\Delta t \lesssim$ 2--3 s and suppression at $\Delta t  =$ 2--8 s (N95).
We have found that the suppression is so pronounced for shots with 
large peaks that the distributions for shots with $p \gtrsim 2.35\langle{p}\rangle$ 
approach log-normal distributions at $\Delta t >$ 2 s (figure 1).
 The transition to the log-normal distribution seems to undergo not so 
gradually, but around 
$p \simeq 2.3\langle{p}\rangle$ ($\simeq 336$ c/31.25 ms), mainly due to 
a sudden drop at $\Delta t =$ 2--6 s. For instance, the occurrence rate 
at $\Delta t =$ 2.25--3.25 s for shots with $p \geq 2.35\langle{p}\rangle$ 
is about half of what is expected from the Poisson distribution 
at longer intervals of $\Delta t \geq 10.25$ s.

	On the other hand, we have confirmed that enhancements 
at $\Delta t \sim 50$ s and $\Delta t \sim 70$ s in figure 1 are real, 
and not due to data 
gaps, by which intervals are completely excluded in the distributions, 
and other artificial effects. 
 These enhancements make the shot occurrence rates, $\lambda$,
defined as frequency $\propto \exp(-\lambda t)$, smaller, and
this and poor statistics make suppression at $\Delta t =$ 2--6 s in the top 
and middle panels in figure 1 less visible, compared with N95.

We may caution that 
it is not yet clear if the distributions are really log-normals
rather than anything else because of the large error bars.
We also point out, however, that the same might be true for the GRB cases.
The difference between a broken exponential function and a log-normal function
is not so clear at $\Delta t > \mu$, as shown in figure 1.
It might be worth reinvestigating the peak interval distributions 
in previous GRB timing studies, while taking account of 
the suppression at short intervals.

Figure 2 depicts the dependence of the log-normal peak positions, $\mu$,
on the selected peak intensities. A significant shift of the peak position,
$\mu$, by the magnitude of the shots is not recognized. 
Fits to histograms binned into 2 s give
$\mu = 7.5^{+1.9}_{-2.3}$ for $p \geq 2.25\langle{p}\rangle$ at $\Delta t \geq 6$ s,
$\mu = 7.5^{+1.2}_{-1.2}$ for $p \geq 2.35\langle{p}\rangle$ at $\Delta t \geq 2$ s,
$\mu = 6.9^{+2.1}_{-2.3}$ for $p \geq 2.5\langle{p}\rangle$ at $\Delta t \geq 2$ s,
and
$\mu = 9.8^{+3.1}_{-3.4}$ for $p \geq 2.6\langle{p}\rangle$ at $\Delta t\geq 2$ s.
This suggests the existence of a particular time interval in this system.

\begin{figure}[t]
   \FigureFile(80mm,60mm){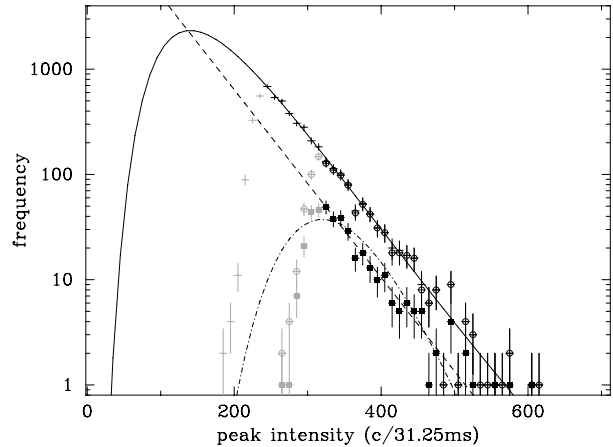}
\caption{
Peak intensity distributions of shots with 
$p \geq 1.5\langle{p}\rangle$ (points with 1 $\sigma$ errors), $p \geq 2.0\langle{p}\rangle$ (open circles),
and $p \geq 2.0\langle{p}\rangle$ and preceding {\it or} following intervals for $\Delta t > 10$ s 
 (closed squares).
Data in incomplete accumulated regions are shown by light grey.
The solid and the dash-dotted line are the best-fit log-normal 
distributions for the 1st and 3rd data, respectively, and the dashed line is 
the best-fit exponential distribution for the 3rd data.
}
\end{figure}

\subsection{Peak Intensity Distribution}
The number of a peak-bin count, $p$, is approximately proportional to the 
total number of counts of the shot, i.e., ``peak fluence'',
since the shot profile is almost independent of the peak intensity 
(Negoro et al. 1994).
To investigate the peak intensity distribution, we only used the 
May 9 and 11 data with similar local mean numbers of counts, 
$\langle{\overline{p}}\rangle \simeq 150$ c/31.25 ms.
We, however, did not exclude data with any $\langle{p}\rangle$, different
from N95, because of the change in the estimation of $\langle{p}\rangle$.
This does not result in any qualitative difference in the following results.

We found that the peak intensity distribution is well consistent with
a log-normal distribution, though no typical peak intensity can
be seen (points with 1 $\sigma$ errors and solid line in figure 3).
Of course, this is not a surprising matter because the shape of the 
right side of a log-normal function is similar to an exponential 
function, as already noted.
A fit to the data at $p > 240$ c/31.25 ms, where the data are almost
completely accumulated, with a log-normal function
gives $\mu = 140\pm14$ c/31.25 ms\footnote{
We could not obtain precise errors, defined as $\Delta \chi^2$ = 2.7,
from the maximum likelihood fits to peak intensity distributions.
The error shown is an approximate 1-$\sigma$ error given by {\tt QDP}.
A $\chi^2$ fit in the above case gives $\mu = 164^{+17}_{-20}$.
}.
 Note that this peak intensity at the maximal frequency is almost the same 
level as the mean count rate $\sim 150$ c/31.25 ms.
The shape of the peak intensity distribution depends slightly 
on $T_{\rm p}$ (N95), but the distribution is always consistent
with a log-normal function with $\mu =$ 100--200 c/31.25 ms.
Thus, the existence of the peak is still a matter to be confirmed.

In the previous subsection, we have shown that large shots 
give rise to the log-normal distributions for the peak intervals. 
Inversely, do shots with long intervals as producing 
a log-normal distribution for time intervals 
have a log-normal distribution for peak intensity?
To investigate this problem, we only selected shots with $p \geq 2.0\langle{p}\rangle$
(open circles),
of which peak intervals are widely distributed up to $\sim 40$ s
(cf., those for shots with $p \geq 1.5\langle{p}\rangle$ are only up to 
$\sim 15$ s; see the top panel of figure 1 in N95).
The data show that all distributions for shots with long preceding 
and/or following 
intervals, say $\Delta t >$ 5 s or 10 s (closed squares), are
more consistent with an exponential distribution (dashed line)
than a log-normal distribution having a typical peak intensity 
at a high count rate (dash-dotted line for the case of 
$\mu \sim 320$ c/31.25 ms).
Thus, a tendency to support the above hypothesis is not recognized.

\section{Discussion}

In the present study we re-examined the temporal varibility
of Cyg X-1 and found for the first time
a similarity to the variability of GRBs.
The similarity and differences should contain important physics
to probe the central engine of GRBs.

The most intriguing difference is that the Cyg X-1 shots, {\it including small ones},
have  smooth (exponential) peak-interval and peak-intensity 
distributions, and 
GRBs and blazar flares, on the other hand, have log-normal distributions,
indicating the presence of a typical timescale and size of energy.
Note that a smooth (power-law) peak-intensity distribution is 
known to exist in solar flares (Dennis 1985), and is also
found in 3D MHD flow simulation data (Kawaguchi et al. 2000;
Agol et al. 2001).
Such a power-law distribution 
is claimed to be ubiquitous in nature (Bak 1996),
and is often discussed in terms of the dynamics of
diffusion systems with interacting degrees of freedom 
(self-organized criticality, see Bak et al. 1988; Mineshige et al. 1994;
arguments about the exponential and power-law distributions, 
see Takeuchi et al. 1995).  
It can be conjectured that MHD turbulence created by
various MHD processes in accretion disks exhibit spatial fractal patterns 
which could be responsible for the smooth distributions of flare amplitudes
(Kawaguchi et al. 2000).

On the other hand, the typical peak interval duration, 7--8 s, 
found in the (log-normal) peak interval distributions 
for {\it large} shots, seems to correspond to the timescales of 
filling/refilling energy in the innermost part of the accretion flow 
before/after the shot occurrence (e.g., N95).
What happens if a large amount of (sudden) accretion takes place
in the inner part of the accretion disk?
VLBI observations of the extragalactic jets 
(Junor et al. 1999)
and recent MHD jet simulations (Kudoh et al. 2002) 
indicate that jets originate from a compact region with a size of
several tens to hundreds of Schwarzschild radii.  
It is of great importance to note that
radio flares are observed only after large X-ray flares
in the Galactic microquasar, GRS 1915$-$105 (Mirabel et al. 1998).
Furthermore, recent discovery of short-term temporal correlation
between X-rays and optical in the BHC, XTE J1118+480, strongly suggests 
that the outflow (or jet) follows the shot (Kanbach et al. 2001).

It may be that jets can be produced only when large enough 
disturbances have been added to the innermost part of the accretion flow.

To summarize, we
newly discovered a similar behavior in Cyg X-1 to that of GRBs 
and blazars, if we only selected large shots from 
the variability light curves of Cyg X-1.
There could also be large- and small-amplitude variations in 
accretion disks at the centers of GRBs, as in GBHCs,
but only large variations (which satisfy a certain criterion) can produce
observable bursts through jets in GRBs.

\bigskip

This work was initiated through a discussion in the domestic GRBs workshop 
held at the Yukawa institute.
HN thanks the Yukawa institute for the hospitality to write
a part of this paper.
This work was supported in part by the Grants-in Aid of the
Ministry of Education, Culture, Sports, Science and Technology of Japan
(13640238, SM).

\end{document}